\documentclass[prx,twocolumn]{revtex4}

\usepackage{amsmath,amssymb} 
\usepackage{bm}
\usepackage[colorlinks=true, linkcolor=blue, urlcolor=blue, citecolor=blue]{hyperref}
\usepackage{subfigure}
\usepackage{graphicx}
\usepackage{color}
\usepackage[dvipsnames]{xcolor}
\usepackage{subfigure}
\usepackage{sidecap}
\usepackage{pgfplots}
\usepackage{tikz}
\usepackage{cleveref}
\usepackage{lettrine}
\write18{-shell-escape}
\usetikzlibrary{external}
\usetikzlibrary{datavisualization}
\usetikzlibrary{matrix}
\usepackage{pgfplots}
	\usepgfplotslibrary{groupplots}
\pgfplotsset{width = 1in, compat=1.3}

\newcommand{\unitv}[1]{\hat{#1}}

\newcommand{\ket}[1]{|\rule{0cm}{2ex} #1 \rangle}

\newcommand{\wsymbol}[6]{ \left\{\begin{array}{ccc} #1 & #2 & #3 \\ #4 & #5 & #6 \end{array} \right\}  }

\newcommand{\zhat}{\unitv{z}}
\newcommand{\ehat}{\unitv{\varepsilon}}
\newcommand{\bhat}{\unitv{b}}
\newcommand{\khat}{\unitv{k}}

\colorlet{Col1}{blue}
\colorlet{Col2}{red}

\colorlet{Col3}{ForestGreen}
\colorlet{Col4}{orange}

\begin{document}

\title{Measurement of the $^{87}$Rb D-line vector tune-out wavelength}

\date{\today}

\author{A. J. Fallon}
\author{E. R. Moan}
\author{E. A. Larson}
\author{C. A. Sackett}
\email[Email: ]{cas8m@virginia.edu}
\affiliation{Department of Physics, University of Virginia, Charlottesville, Virginia 22904, USA}

\begin{abstract}
We report a precision measurement of a tune-out wavelength for $^{87}$Rb using circularly polarized light. A tune-out wavelength characterizes a zero in the electric polarizability of the atom. For circularly polarized light, the total polarizability depends on both the scalar and vector polarizability components. This shifts the location of the tune-out wavelength and makes it sensitive to different combinations of atomic dipole matrix elements than the scalar polarizability alone. Using $\sigma_-$ polarized light with an an estimated purity of 0.9931(1), we observe a tune-out wavelength of 785.1522(3)~nm, which agrees with theoretical expectations when small contributions from the core electrons and off-resonant valence states are taken into account. 
\end{abstract}

\maketitle
 
 A tune-out wavelength describes a light frequency at which an atom or molecule in a given state experiences zero first-order energy shift from an optical field \cite{LeBlanc2007,Arora2011}. Tune-out wavelengths find applications in experiments involving multiple species, where it can be useful to apply an energy shift to one species without affecting another \cite{Lamporesi2010,Clark2015}. Tune-out wavelength measurements are also useful in their own right because they provide information about the dipole matrix elements of the target particle that may not otherwise be easily accessible. Knowledge of dipole matrix elements is important for many reasons, including the interpretation of parity violation experiments, accurate estimation of black-body radiation shifts in atomic clocks \cite{Mitroy2010,Safronova2012}, and as benchmarks for atomic theory calculations. These benefits have prompted a series of precise tune-out-wavelength measurements in alkali and other atoms \cite{Lamporesi2010,Herold2012,Holmgren2012,Jiang2013,Henson2015,Leonard2015,Trubko2015,Schmidt2016,Copenhaver2019,Decamps2020,Wen2021}. These experiments have mainly focused on zeros of the scalar electric polarizability of the atoms. However, additional information can be obtained from the vector character of the polarizability, which is exhibited through a dependence on the optical polarization of the applied light \cite{Trubko2015,Schmidt2016,Wen2021}. We here explore this polarization dependence through a precise measurement of a vector tune-out wavelength.
 
 Vector tune-out measurements are useful both for trapping applications and for fundamental atomic physics. For applications, they add flexibility by allowing the tune-out value to be adjusted \cite{Clark2015}. For instance, the D-line scalar tune-out wavelength for Rb is fixed at 790.032 nm, but by adjusting the light polarization, the tune-out wavelength can be set anywhere between 785.112 nm and the D1 line at 794.978 nm. This flexibility can make it easier to satisfy other experimental requirements. It can also be useful that vector fields cause the tune-out wavelength to depend on the magnetic sublevel of the particle \cite{Schneeweiss2014,Trubko2015,Schmidt2016}. 
 
 In regards to fundamental physics, precise measurements of the polarization allows resolution of contributions to the atomic polarizability from different angular momentum states. For instance, interpreting alkali atom parity violation amplitudes in terms of nuclear physics parameters requires knowledge of the $nS_{1/2} \leftrightarrow nP_{1/2}$ dipole matrix elements \cite{Porsev2010,Dzuba2012b}. Vector tune-out measurements can allow the $P_{1/2}$ matrix elements to be constrained separately from the $P_{3/2}$ matrix elements, whereas a purely scalar measurement depends jointly on both $P_{1/2}$ and $P_{3/2}$ elements \cite{Fallon2016}.
   
 The theoretical framework for vector tune-out wavelengths is well understood \cite{Manakov1986,Sahoo2013,Kien2013,Wang2020}. However, making a precise comparison between theory and measurement requires careful control of both the light polarization and the alignment of the laser beam axis to the quantizing magnetic field. The measurement reported here has a wavelength precision of order 1 pm, and agrees to this level with theoretical expectations. At this precision, the measurement is sensitive to small effects including the polarizability of the ionic core and contributions from far off-resonant valence states. With realistic improvements in precision and by combining tune-out measurements for different states, the technique could provide constraints on important dipole matrix elements and yield accuracies better than the best current theoretical uncertainties.
 
 The energy shift of a particle in an optical field $\mathcal{E}$ can be expressed as
\begin{equation}
    U = -\frac{1}{2}\alpha \langle \mathcal{E}^2\rangle = -\frac{1}{2\epsilon_0 c} \alpha I,
\end{equation}
where $\alpha$ is the electric polarizability, $I$ is the light intensity, $c$ is the speed of light, and $\epsilon_0$ is the electric constant. The polarizability depends on both the frequency $\omega$ and the polarization state $\ehat$ of the light. For an atom in ground hyperfine state $\ket{n,J,F,m}$, a spherical tensor decomposition gives \cite{Sahoo2013,Kien2013}
\begin{align}
    \alpha & = \alpha^{(0)} - \alpha^{(1)} S_3 \unitv{k}\cdot\unitv{b} \frac{m}{2F}  \nonumber \\
        & + \alpha^{(2)} \left[\frac{3|\ehat\cdot\unitv{b}|^2 -1}{2}\right] \frac{3m^2-F(F+1)}{F(2F-1)}  ,
\end{align}
where the $\alpha^{(i)}$ parameters are the scalar, vector, and tensor polarizability components, for $i = 0, 1,$ and 2, respectively. Here the light field is taken as a plane wave propagating in direction $\unitv{k}$, with complex polarization vector $\ehat$. The atomic states are defined relative to a magnetic field pointing in direction $\unitv{b}$. The parameter $S_3 = i(\ehat^*\times\ehat)\cdot\unitv{k}$ is the fourth Stokes parameter for the light, with $S_3 = \pm 1$ for $\sigma_\mp$ circularly polarized light. Our measurements use the $5S_{1/2}$ ground state of $^{87}$Rb, with $F = m = 2$.

In the case of an alkali atom, the polarizability components can be separated into a contribution from the valence electron, a contribution from the core electrons, and a term reflecting interactions between the valence electron and the core. The valence contribution can be calculated using perturbation theory as a sum over excited $P$ states $\ket{n',J',F',m'}$. We measure the tune-out wavelength near the $5P_{1/2}$ and $5P_{3/2}$ states, so an accurate calculation needs to account for the hyperfine splittings of these states. For higher-lying states, hyperfine shifts can be neglected. We therefore express the polarizability as
\begin{equation}
    \alpha^{(i)} = \alpha_c^{(i)} + \alpha_{cv}^{(i)} + \alpha_{5P}^{(i)} + \alpha_{v'}^{(i)}
\end{equation}
where $\alpha_c$ denotes the core contribution, $\alpha_{cv}$ the core-valence interaction, $\alpha_{5P}$ the contribution from the $5P$ states,
and $\alpha_{v'}$ the contribution from other valence states. Furthermore, the core contribution has only a scalar component $i = 0$, since the core is spherically symmetric. We also neglect the tensor components $\alpha_{cv}^{(2)}$ and $\alpha_{v'}^{(2)}$ since they are very small. The remaining valence contributions are then \cite{Kien2013}:
\begin{align}
    \alpha_{5P}^{(0)} & = \frac{2}{\hbar}\frac{1}{\sqrt{3(F+1)}} \sum_{J',F'} \frac{|d'|^2 \omega'}{\omega'^2-\omega^2} 
    \wsymbol{1}{0}{1}{F}{F'}{F} C'\\
    \alpha_{5P}^{(1)} & = \frac{2}{\hbar}\sqrt{\frac{2F}{(F+1)(2F+1)}}  \nonumber \\
    & \qquad \times \sum_{J',F'}\frac{|d'|^2 \omega}{\omega'^2-\omega^2} 
    \wsymbol{1}{1}{1}{F}{F'}{F} C'\\
    \alpha_{5P}^{(2)} & =   \frac{2}{\hbar}\sqrt{\frac{2F(2F-1)}{3(F+1)(2F+1)(2F+3)}}
    \nonumber \\
    & \qquad \times \sum_{J',F'} \frac{|d'|^2 \omega'}{\omega'^2-\omega^2} 
    \wsymbol{1}{2}{1}{F}{F'}{F} C'\\
    \alpha_{v'}^{(0)} & =  \frac{1}{3\hbar} \sum_{n',J'} \frac{|d'|^2 \omega'}{\omega'^2-\omega^2} \label{alphav0} \\
    \alpha_{v'}^{(1)} & =  \frac{1}{3\hbar} \sum_{n',J'} \frac{|d'|^2 \omega}{\omega'^2-\omega^2} \left(3J'-\frac{7}{2}\right) \label{alphav1}
\end{align}
Here $\omega$ is the light frequency, and $\omega'$ is the transition frequency from $\ket{5S_{1/2},F}$ to $\ket{n'P_{J'},F'}$. We neglect Zeeman shifts since they are small compared to our measurement precision. The arrays in braces are Wigner $6-j$ symbols. The reduced matrix elements are 
$d' \equiv \langle 5S_{1/2} || d || n'P_{J'}\rangle$, and in the $\alpha_{5P}$ terms we use
\begin{equation}
    C' = (-1)^{F+F'+1} (2F+1)(2F'+1)\wsymbol{F}{1}{F'}{J'}{I}{J}^2
\end{equation}
with nuclear angular momentum $I = 3/2$.  

\begin{figure}
\includegraphics[width=3.3in]{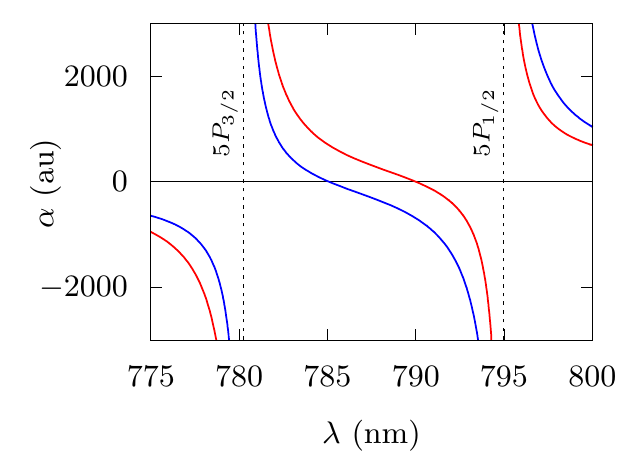}
\caption{(color online) Electric polarizability of $^{87}$Rb in the $F = 2$, $m = 2$ ground state, as a function of optical wavelength. The red curve shows the case of linearly polarized light, and exhibits a tune-out wavelength near 790~nm. The blue curve shows the case of $\sigma_-$ polarized light, with a tune-out wavelength near 785~nm. \label{fig:alpha}}
\end{figure}

We take $\alpha_c^{(0)} = 9.116(9)$ from Ref.~\cite{Berl2020}, and $\alpha_{cv}^{(0)} = -0.37(4)$ and $\alpha_{cv}^{(1)} = -0.04(4)$ from Ref.~\cite{Fallon2016}. For the $5P$ states we use $d_{5P_{1/2}} = 4.234(2)$ and the ratio $d_{5P_{3/2}}/d_{5P_{1/2}} = 1.99217(3)$ from \cite{Leonard2015}. For higher-lying valence states we use the matrix elements tabulated in \cite{Leonard2015}. With these values, we can calculate the net polarizability $\alpha$ for given values of the experimental parameters $S_3$, $\khat\cdot\bhat$, and $\ehat\cdot\bhat$. 
Figure \ref{fig:alpha} shows how $\alpha$ varies with wavelength for the cases of linear and $\sigma_-$ polarized light. The tune-out wavelength is located where $\alpha = 0$. 
Table \ref{tab:alpha} lists the various contributions to $\alpha$ at the tune-out wavelength for $\sigma_-$ polarized light.

\begin{table}
\bgroup
\def\arraystretch{1.5}
\setlength\tabcolsep{2ex}
\begin{tabular}{cc|cc}
Term & Value (au) & Term & Value (au)\\ \hline
$\alpha_{5P}^{(0)}$ & 12347.7(4)(11.6)  & $\alpha_{v'}^{(1)}$ & 0.2(1) \\
$\alpha_{5P}^{(1)}$ & 24716.8(4)(23.3)  & $\alpha_c^{(0)}$ & 9.12(1) \\
$\alpha_{5P}^{(2)}$ & -0.04352(1)(4)  & $\alpha_{cv}^{(0)}$ & -0.37(4) \\
$\alpha_{v'}^{(0)}$ & 2.0(1) & $\alpha_{cv}^{(1)}$ & -0.04(4) 
\end{tabular}
\egroup
\caption{Contributions to the total polarizability $\alpha$ at the tune-out wavelength $\lambda = 785.112$ nm for ideal $\sigma_-$ polarized light with $S_3 = \khat\cdot\bhat = 1$, $\ehat\cdot\bhat = 0$. Values in parentheses show the estimated errors. In the case of the $\alpha_{5P}$ contributions, the first parentheses show the uncertainty arising from the uncertainty in the ratio of the $d_{5P1/2}$ to $d_{5P3/2}$ matrix elements. The second parentheses show the uncertainty from the $d_{5P1/2}$ element itself, which is large but correlated among the different components, and therefore has negligible impact on the value of the tune-out wavelength.
\label{tab:alpha}
}
\end{table}

The experimental apparatus is similar to that used in Ref.~\cite{Leonard2015}. A Bose-Einstein condensate of about $10^4$ atoms is produced in a weak magnetic trap, with harmonic oscillation frequencies of 5.1, 1.1 and 3.2 Hz along the $x$, $y$ and $z$ directions respectively. The $z$ direction is vertical. The trap uses the time-orbiting potential (TOP) technique, with a bias field of 21.4 G rotating in the $xz$ plane at frequency $\Omega = 2\pi \times 12.8$~kHz. Confinement and support against gravity are provided by oscillating magnetic gradients. 

The tune-out measurement was performed using an atom interferometer. An off-resonant standing-wave laser along the $y$ axis of the trap applied velocity kicks in units of $v_B = 2\hbar k/m = 11.8$~mm/s, via Bragg scattering. The interferometer used a total of four Bragg pulses. At time $t = 0$, an initial pulse split the condensate into wave packets moving at $\pm v_B$. At time $t = 10$~ms, the laser was applied again so as to reverse the atoms' motion. The packets then passed through each other with minimal interactions, and at $t = 30$~ms a third laser pulse reversed the motion again. Finally at $t = 40$~ms, the initial splitting pulse was reapplied and the wave packets were recombined. A fraction $N_0/N$ of the atoms were brought back to rest in the center of the trap, with signal
\begin{equation} \label{eq:signal}
S = \frac{N_0}{N} = \frac{1}{2}[1+V \cos(\phi+\phi_r)].
\end{equation}
Here the wave packets have developed a phase difference $\phi$, the phase of the recombination pulse relative to the initial splitting pulse is $\phi_r$, and the visibility is $V = 0.7$. The fraction of atoms at rest was detected by absorption imaging after a short time of flight. We set $\phi_r \approx \pi/2$ by shifting the frequency of the Bragg laser prior to the final pulse.

A Stark phase shift $\phi$ was applied by directing a second laser beam, traveling along $z$, onto one arm of the interferometer. The beam was focused to a waist of about 50 $\mu$m, which is smaller than the maximum wave-packet separation of 240 $\mu$m. The Stark beam was derived from an MBR Ti:Sapphire laser. This is an improvement over the tapered amplifier used in our previous work \cite{Leonard2015}, since the Ti:Sapphire laser is not expected to contain a significant amount of amplified spontaneous emission light at other frequencies. The Stark beam was applied for 20 ms at the start of the interferometer, so that one packet passed through it twice. This leads to a phase
$\phi = \int \alpha I\,dt/(2\epsilon_0 \hbar c)  $
which the interferometer detects. 

To control the vector portion of the polarizability, we used a pair of acousto-optic modulators to pulse the Stark beam synchronously with the rotating bias field, such that the light was on only when the field pointed along $z$. Two modulators were used to provide an extinction ratio better than 60~dB. The Stark beam was aligned to the field by tuning to the D1 resonance at 795~nm and setting the polarization state to $\sigma_+$. When the laser is optimally aligned to the magnetic field, the atoms scatter no light since there is no $m=3$ state in the D1 hyperfine manifold. Details of this measurement are provided in Ref.~\cite{Fallon2020}. From the residual scattering rate, the alignment error $\delta\theta$ between $\khat$ and $\zhat$ was constrained to be less than 16 mrad. In the interferometry experiments, the duration of the pulses was $\tau = 5~\mu$s, and the angle between $\khat$ and $\bhat$ varied during the pulse as the field rotated at frequency $\Omega$. This gives an average value for $\khat\cdot\bhat$ of $(2/\Omega\tau) \sin(\Omega\tau/2) \cos\delta\theta \approx 0.99321(6)$, where the uncertainty reflects the angular misalignment. Other sources of uncertainty, such as variations in the pulse length and non-uniformity of the pulse, are about an order of magnitude smaller. The integrated value of $\ehat\cdot\bhat$ in the tensor term can similarly be calculated as $\Omega^2\tau^2/48 \approx 3.3\times 10^{-3}$, but this is insignificant because the tensor component $\alpha^{(2)}$ is much smaller than the vector component $\alpha^{(1)}$.

It is also critical to control the light polarization accurately. We set the polarization close to circular using a calcite polarizer, two wave plates, and a Fresnel rhomb, as described in \cite{Fallon2020}. The wave plates provide a small correction to the Fresnel rhomb to account for polarization distortions in the vacuum window and other optics. The polarization can be set accurately using again the photon scattering measurements at 795 nm, but the mirrors that direct the beam onto the atoms are slightly chromatic and the polarization is not sufficiently preserved when the Stark laser is scanned to the tune-out wavelength at 785 nm. Instead, we set the laser slightly blue of the tune-out wavelength, and adjusted the wave plates to minimize the interferometer phase $\phi$. Since the tune-out wavelength is as blue as possible for $\sigma_-$ polarized light, this optimizes the polarization at the atoms. The wave plate angles could be optimized to an accuracy of about 0.5$^\circ$, from which we estimate $S_3 = 0.99988(12)$.

\begin{figure}
\includegraphics[width=3.3in]{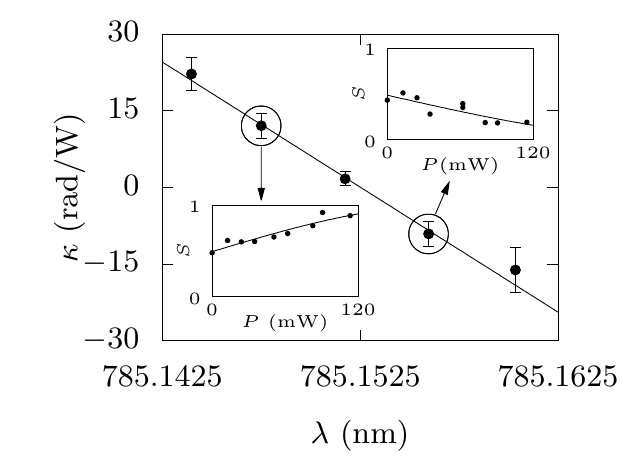}
\caption{Tune-out measurement data. For each of the points in the main graph, an interferometric measurement is performed to determine $\kappa = \phi/P$, where $\phi$ is the interferometer phase and $P$ is the peak power of the Stark beam pulses. The inset graphs show example plots of the interferometer signal vs.\ power, from which $\kappa$ is determined via a fit to the form of Eq.~\protect\eqref{eq:signal}. The $\kappa$ data in the main graph are fit to a line, and the $x$-intercept of 785.1525 nm is taken as the tune-out wavelength value. \label{fig:data} }
\end{figure}

To perform the measurement, we set the Stark laser to a series of wavelengths near 785 nm, and at each wavelength we varied the pulse power $P$ to scan $\phi$ and trace out an interference curve. Example data are shown as insets in Fig.~\ref{fig:data}. We assume $\phi = \kappa P$ and fit the trace data to obtain a value for $\kappa$. The main graph of Fig.~\ref{fig:data} shows that near 785 nm, $\kappa$ is a linear function of wavelength which crosses zero at 785.1525(5)~nm. Here the uncertainty is primarily from noise in the linear fit, but also includes the 0.15~nm uncertainty in our wave meter calibration. A second independent measurement yielded a consistent value of 785.1519(4) nm, so we report the average of these results as $\lambda_0 = 785.1522(3)$ nm.

In comparison, using the the experimental estimates for $\khat\cdot\bhat$ and $S_3$, we calculate an expected tune-out value of 785.1538(9)~nm, which is about $2\sigma$ different from our measurement. The optical polarization is the largest source of uncertainty in the theoretical value, with the alignment error and atomic parameters contributing about half as much. Table \ref{tab:errors} summarizes the main contributions to the uncertainties of the measurement and calculation. Our result is consistent with that obtained by Wen {\em et al.} \cite{Wen2021}, who found $\lambda_0 = 785.146(12)$ nm for $\sigma_-$ polarized light.

Although the discrepancy between our measurement and calculation is not large enough to be significant, the sign is interesting, since larger-than-estimated errors in the polarization or alignment would result in a measurement redder than expected, whereas our result is bluer. If we assume that the Stark beam parameters are perfect and account only for the rotation of the bias field during the Stark pulses, we would expect a tune-out wavelength of 785.1530(4) nm, still about $2\sigma$ redder than observed. 

\begin{table}
\begin{tabular}{lcc}
Source & $\delta\lambda_0$ (pm) & $\delta\alpha$ (au) \\ \hline
Measurement total: & 0.46 & 1.2 \\
\hspace{1em} Statistical & 0.43 & 1.1 \\
\hspace{1em} Wavemeter & 0.15 & 0.4 \\
Calculation total: & 0.88 & 2.2 \\
\hspace{1em} Atomic parameters & 0.35 & 0.9 \\
\hspace{1em} Polarization & 0.70 & 1.8 \\
\hspace{1em} Alignment & 0.35 & 0.9 \\
\hspace{1em} Pulse length & 0.06 & 0.2 \\
\hspace{1em} Pulse symmetry & 0.06 & 0.2 
\end{tabular}
\caption{Sources of error in the tune-out wavelength measurement and calculation. For each contribution, the impact is given both as the uncertainty $\delta\lambda_0$ in the tune-out wavelength and as the uncertainty in the value $\delta\alpha$ of the polarizability at the measured tune-out wavelength. These are related using the calculated derivative $|d\alpha/d\lambda| = 2.527$~au/pm. The uncertainty contribution labeled as `atomic parameters' refers to the values in Table~\protect\ref{tab:alpha}. \label{tab:errors}} 
\end{table} 
 
For atom-trapping applications, the level of precision demonstrated here shows how accurately a vector tune-out application can be implemented. For instance, in the conditions of our experiment, a rubidium atom in the $F = 2, m = 1$ state would experience a total polarizability $\alpha \approx 6222$ au. This can be compared to a residual polarizability $\delta\alpha \approx 2$ au for an $F = 2, m = 2$ atom. The ratio $\alpha/\delta\alpha = 3\times10^3$ indicates how strongly the $m = 1$ atom can be manipulated before the $m = 2$ atom is affected. 

In terms of atomic physics, we see that the precision demonstrated here is already sufficient to distinguish the larger non-$5P$ contributions to the net polarizability. If, for instance, the core contribution $\alpha_c^{(0)}+\alpha_{cv}^{(0)}$ were excluded from the calculation, the expected tune-out wavelength would shift by about $4\sigma$. Our measurement thus tests the theory in a non-trivial way, but with reasonable increases in precision, it could provide a more meaningful comparison. For instance, with a factor of five improvement the experiment would be sensitive to the $\alpha_{cv}^{(0)}$ core-valence interaction, which has not previously been experimentally observed. With a factor of fifty improvement, the experimental precision would exceed the theoretical precision in most cases. This would be particularly interesting for the $\alpha_{v'}$ terms, where the theoretical uncertainty is dominated by the contribution from the high $n'$ Rydberg tail. This same contribution is the largest source of uncertainty in the relationship between measured atomic parity violation amplitudes and the weak mixing angle of the standard model \cite{Dzuba2012b}, so providing a precise benchmark via the polarizability can be expected to help improve the parity violation interpretation.

Improvement by a factor of fifty is experimentally feasible. We have previously demonstrated a scalar tune-out measurement with an uncertainty of 0.035 pm, which was limited primarily by statistics \cite{Leonard2015}. Improvement to 0.01 pm should involve no new challenges. The vector measurement is more difficult due to the requirement for polarization control, but many of the limitations encountered here could be resolved for atoms confined in an optical trap rather than a TOP trap, since it would then be possible to use a static bias field and a continuous-wave Stark beam. This would allow significantly higher average power to be applied to the atoms, so that the polarization and alignment optimizations could be made more precise. Recent experiments at Los Alamos National Laboratory have demonstrated an optically trapped atom interferometer with performance comparable to that used here \cite{Boshier2021}. We therefore argue that reaching experimental precision comparable to the theoretical precision is likely achievable. 

Looking forward to such experiments, it will be necessary to distinguish the various contributions to $\alpha$ so that, for instance, the Rydberg tail contribution can be isolated from the core-valence interaction. This can be achieved by comparing tune-out measurements near different states, such as the $6P$ states near $\lambda = 420$~nm for Rb. The core and Rydberg contributions have different frequency dependencies, allowing their impact to be resolved \cite{Fallon2016}. Further, since the $J=1/2$ and $J=3/2$ states contribute differently to the scalar and vector components of $\alpha_{v'}$ in Eqs.~\eqref{alphav0} and \eqref{alphav1}, these two contributions can be distinguished as well. The parity violation interpretation depends only on the $J=1/2$ matrix elements. We therefore expect vector measurements to be an important component of this approach.

In summary, we have carried out a precise measurement of a vector tune-out wavelength, for near-circularly-polarized light. We show that the polarization and alignment factors can be controlled with 10 ppm precision, even in the rotating magnetic field of a TOP trap. The 1 ppm precision that we obtain in the wavelength is similar to that of many scalar tune-out measurements, but the vector character provides both more utility and more information. We believe that this work illustrates the feasibility and utility of precise vector tune-out measurements, and we hope that our results stimulate further improvements to the point that the method becomes useful for interpreting parity violation and other experiments that rely on atomic dipole matrix elements.

\begin{acknowledgments}
This work was supported by the National Science Foundation (Grant Nos.~PHY-1607571 and PHY-2110471) and NASA (Contract No.~RSA1640951). We thank Seth Berl for contributions to the experimental apparatus.
\end{acknowledgments}

\bibliographystyle{apsrev}
\bibliography{Fallon}

\end{document}